\documentclass{PoS}

\title{The FAMU experiment at RIKEN RAL for a precise measure of the
proton radius}

\ShortTitle{The FAMU experiment at RIKEN RAL}

\author{\speaker{M. Bonesini}\thanks{{\bf on behalf of the FAMU Collaboration}}\\
        Sezione INFN Milano Bicocca, Dipartimento di Fisica G. Occhialini, \\
Universit\`a di Milano Bicocca, Italy\\
        E-mail: \email{maurizio.bonesini@mib.infn.it}}

\abstract{
The goal of the FAMU experiment at RIKEN RAL is the measure of the hyperfine 
splitting  of the ground state of the muonic hydrogen, to allow a determination
of the proton Zemach radius with a precision better than $5 \times 10^{-3}$ . The 
comparison of
this measurement with the ones done with ordinary hydrogen may help to solve
the so-called ``proton radius puzzle'' , triggered by the $6 \sigma$ discrepancy
in the proton charge radius value as extracted from muonic Lamb shift and 
the  value  based
on e-p scattering  and ordinary hydrogen spectroscopy.
          }

\FullConference{
European Physical Society Conference on High Energy Physics - EPS-HEP2019 -\\
			10-17 July, 2019\\
			Ghent, Belgium}

\begin{document}

\section{Introduction}
Many properties of the proton, 
such as its radius and anomalous magnetic moment, are not completely
 understood.
The so-called proton radius ``puzzle'' \cite{Antognini} refers to the $6 \sigma$ discrepancy
between the electron and muon determination of the proton charge radius.
This discrepancy may be due to a violation of the electron-muon universality
or simply to not well understood experimental systematic errors.
%

The FAMU (\underline{F}isica degli \underline{A}tomi \underline{Mu}onici) 
experiment aims at the  measurement of  $\Delta E^{hfs}(\mu^-p)_{1S}$ 
with a relative accuracy better than 
$10^{-5}$ \cite{6bakalov,7bakalov,8adamczak}.
It makes
use of a high intensity pulsed low-energy muon beam, stopping in a hydrogen
target, to produce muonic hydrogen (in a mixture of singlet F=0 and triplet
F=1 states) and a tunable mid-IR (MIR) pulsed high power laser
to excite the hyperfine splitting (HFS) transition of the 1S muonic
hydrogen (from F=0 to F=1 states).
Exploiting the muon transfer from muonic hydrogen to another
higher-Z gas in the target (such as $O_2$ or Ar),
the $(\mu^{-}p)_{1S}$ HFS transition will be recognized by
an increase of the number of
 X-rays from the $(\mu Z^{*})$ cascade, while tuning the laser
frequency $\nu_{0}$ ($\Delta E_{HFS}=h \nu_{0}$).
From the measure of $\Delta E^{hfs}(\mu^{-}p)_{1S}$  the Zemach radius $r_Z$ of the proton 
 may be deduced with a precision up to $5 \times 10^{-3}$,
thus  sheding  new light on the problem of
the proton radius puzzle.

The FAMU experiment is being  performed in steps, starting from the study of
the transfer rate from muonic hydrogen to another higher-Z gas 
and ending with the  full working setup 
 with the pump MIR laser and a multipass optical cavity to
determine the proton Zemach radius. The preliminary steps have allowed to
determine the best mixture to be used inside the cryogenic target and
optimize the operating conditions.

\section{Experimental layout at RIKEN-RAL}

\begin{figure*}
  \includegraphics[width=0.55\textwidth]{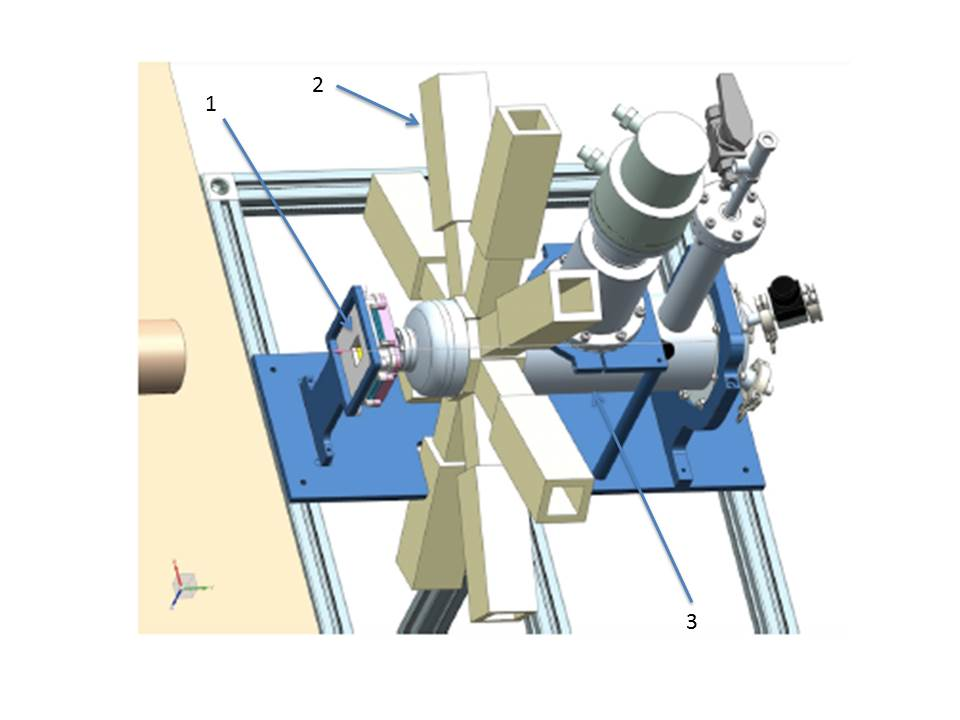}
  \includegraphics[width=0.44\textwidth]{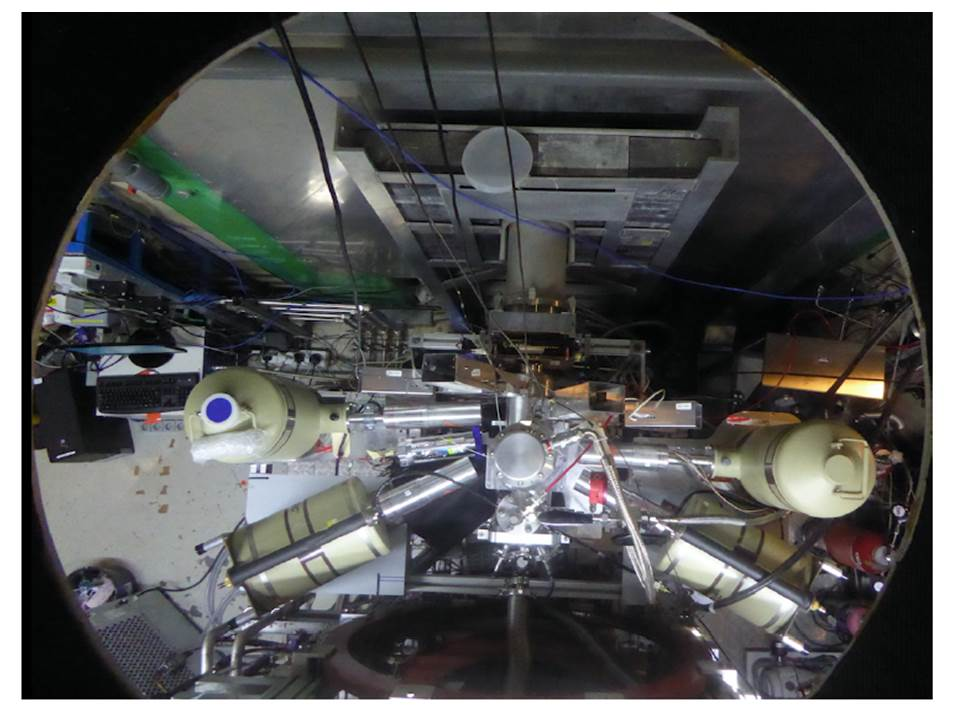}
\caption{Left: layout used in the preliminary data taking of  FAMU  at 
RIKEN-RAL (R582). 1) is the 1 mm pitch beam hodoscope, 2) the crown of 
eight LaBr$_3$:Ce detectors and 
3) the cryogenic target. Right: picture of the experimental setup from the top.
The four HpGe detectors are visible here.}
\label{setup}
\end{figure*}

The setup for the preliminary  2015-2016 (R582)
FAMU data taking is schematically shown
in figure \ref{setup} and fully described in reference \cite{famu2018}. 
It includes a beam hodoscope to characterize the 
impinging muon beam, a cryogenic target  and  detectors to see  the characteristic 
low energy X-rays  emitted in the process under study. 

Data have been taken at the RIKEN-RAL muon facility\cite{45matsuzaki}, where the ISIS primary proton
beam at 800 MeV/c produces a high-intensity pulsed muon beam, impinging on
a secondary carbon target. The experiment makes use of a decay muon beam at $\sim 60$ MeV/c.
The beam has a double pulse structure 
(70 ns FWHM widths, 320 ns peak to peak  distance
) reflecting the primary beam structure. 
With an intensity of $\sim 8 \times 10^4 \mu^{-}/s $ the beam has a 
momentum spread $\sigma_p/p = 4 \%$ and an angular divergence 
$\sim 60$ mrad. 
\subsection{Beam characterization}

\begin{figure*}
\resizebox{0.5\textwidth}{!}{
  \includegraphics{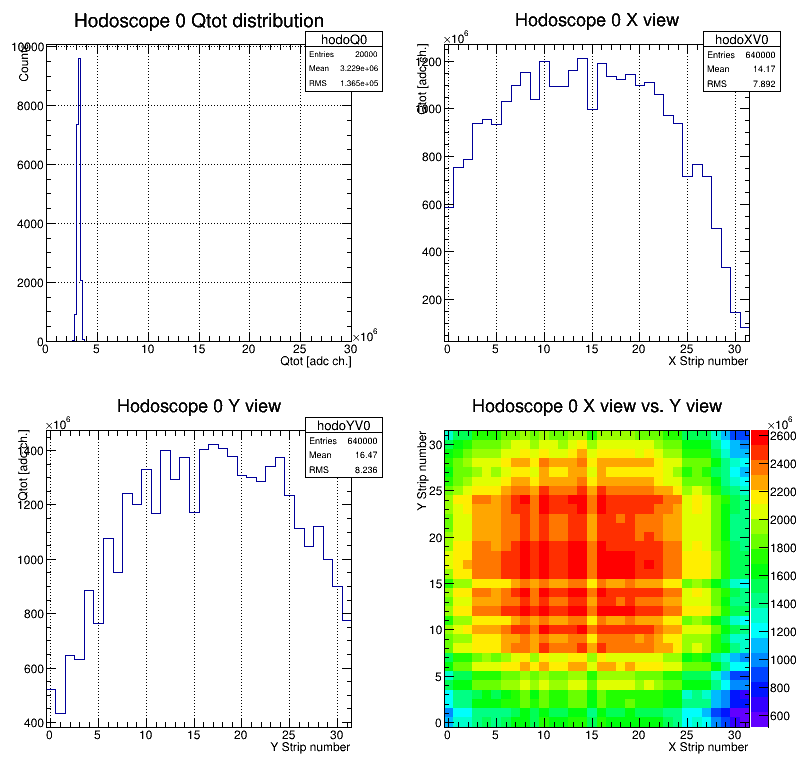}}
\resizebox{0.4\textwidth}{!}{
  \includegraphics{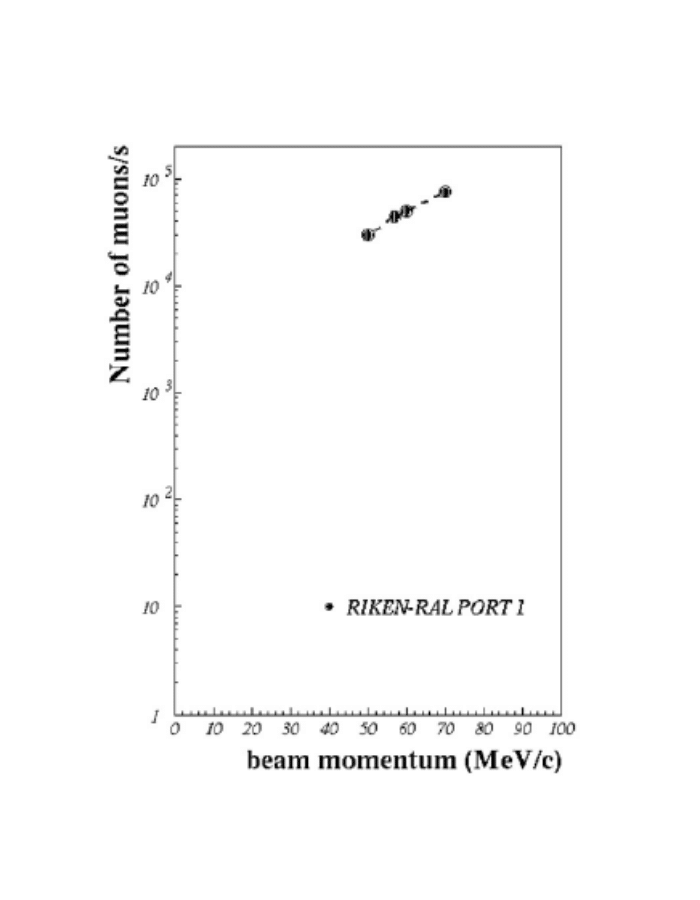}
  }
\caption{Left: from top to bottom, total deposited charge ($Q_{tot}$), X/Y beam projections 
and X/Y beam profile for a 55~MeV/c beam momentum run. Right: beam intensity 
measured as a function of
beam momentum. }
\label{hodoscope}
\end{figure*}
To monitor  the beam parameters,
avoid any undetected drift in beam shape and position and center the
target a beam hodoscope system is needed. 
The beam hodoscope system has allowed to: tune the incoming beam, deliver timing information, as respect to radiofrequency, for DAQ readout and trigger, and monitor the intensity of each beam pulse. The system is based on three hodoscopes. 
Two of them are removable X/Y beam hodoscopes, made  
of 2 planes of 32 3$\times$3 mm$^2$ square scintillating Bicron BCF12 fibers 
read at alternating fibers' ends by Hamamatsu S12572 silicon photomultiplier (SiPM), 
with 25 $\mu$m cells 
\cite{49carbone,49bonesini}. 
They have a 10$\times$10 cm$^2$ active area.
The third one
makes use of  1 mm square BCF12 fibers, with white EMA coating, to
ensure light tightness,   read by 1 mm RGB Advansid SiPM, 
with 40~$\mu$m cells \cite{49bonesini,50bonesini} and has 
an active area of 3.2 $\times$ 3.2~cm$^2$. It was permanently installed 
in front of the beam target window during  data taking. 
As SiPM output signals are sizeable - $\sim$40~mV with a S/B $\sim$10 - no
amplification stage is needed. They are digitized by CAEN V1742 FADCs (5~Gb/s, 12 bits, 1~Vpp dynamic range), providing information on timing, integrated charge and pulse height.
The  gain drift of SiPM with temperature will be controlled by CAEN DT5485 
digital power supplies with built-in feedback on temperature,  
measured by Analog Devices TMP37 thermistors with a $\pm 2 \%$ precision.
A X/Y beam profile, as measured at Port 1, is shown in the left  panel of figure~\ref{hodoscope}. A rms beam width less than 10 mm in both directions (X,Y) is measured. 
The total collected charge from the two planes of a detector ($Q_{tot}$) 
may be used, after a suitable calibration~\footnote{the calibration procedure 
makes use of
laboratory test data taken with cosmic muons and tabulated values of dE/dx vs 
momentum from PDG}, to estimate the incoming muon rate, as shown in the 
right panel of figure~\ref{hodoscope}. \\
To reduce the amount of material in front of the target a
 new hodoscope, based on 0.5 $\times$ 0.5 mm$^2$ square Bicron BCF12 
scintillating fibers 
read by 1$\times$1 mm$^2$ Hamamatsu S12751-050P SiPM will replace the current 
1 mm pitch one in the final FAMU data taking. With 32+32 X,Y channels it will cover an active are of about 7.2 $\times$ 7.2 cm$^2$, to match the area of the new target entrance window.

\subsection{X-rays detection}

To  extract the muonic X-ray lines of interest ( at $\sim$  100-200 keV) 
with the best energy resolution and minimal events pile-up, hygroscopic
LaBr$_3$:Ce crystals were used~\footnote{
A preliminary study  to asses if  non-hygroscopic crystals, such as PrLuAg and
Ce:GAAG, may be suitable was done and  
is reported in reference \cite{bonesini15}}.
In addition HpGe detectors (up to four were used at the same time) completed
the experimental setup. They were used  as a cross-reference due to their much better energy 
resolution. Their readout was performed through CAEN V1724  digitizers 
(14 bit, 100 MS/s, 10 Vpp dynamic range).
Detectors were  placed around the region of maximal density of muonic hydrogen formation.

\begin{figure*}
\begin{center}\resizebox{\textwidth}{!}{
  \includegraphics{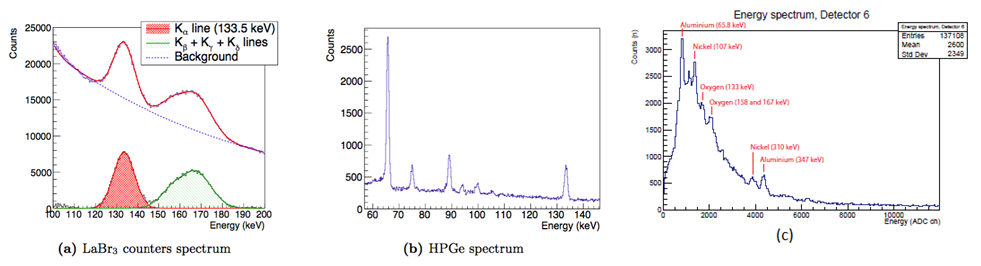}
  }
\end{center}
\caption{Muonic X-ray spectra recorded using (a) 1"
LaBr3:Ce counters, $K_{\beta}$ and $K_\gamma$  lines are not resolved; (b) the HpGe detector and (c) 1/2"  LaBr3:Ce counters with SiPM array readout. }
\label{hpge}
\end{figure*}

The main X-rays detector 
is based on eight one inch diameter, one inch length
 LaBr$_3$:Ce crystals arranged in a crown, read by 
Hamamatsu R112265-200 UBA 1" photomultipliers tube equipped with 
an  active high voltage divider and 
a Digital Pulse Processor (DPP) \cite{baldazzi}. Up to now, data have been acquired and processed by CAEN V1730 digiters (14 bit, 500 MS/s, 2 Vpp dynamic
range) in the framework of the general FAMU Data Acquisition System \cite{52soldani}.

The collaboration has also developed a complementary read-out system for 
the LaBr3:Ce crystals based on SiPM arrays. 
The advantage of the  PMT readout as respect to SiPM readout  of a better rise 
time ($\sim$10 ns as compared to 20 ns) is balanced, as regards the noise, 
by  the use of smaller crystal size: 1/2" instead of 1".
Eight 1/2" LaBr with SiPM array readout have been recently tested and were used,
due to their compact size,  to instrument the most inaccessible regions around the 
target\cite{famu2018,56bonesini, 51benocci}. 
The readout was based on 4$\times$4 array of 3$\times$3 mm$^2$ Hamamatsu S13361 TSV SiPM. The output signals are summed up on a custom made PCB and digitized by CAEN V1730 digitizers. No amplification is needed, since the signal is about 100-200~mV.

\section{Experimental results on the transfer function}

\begin{figure}
\begin{center}
\resizebox{0.8\textwidth}{!}{
  \includegraphics{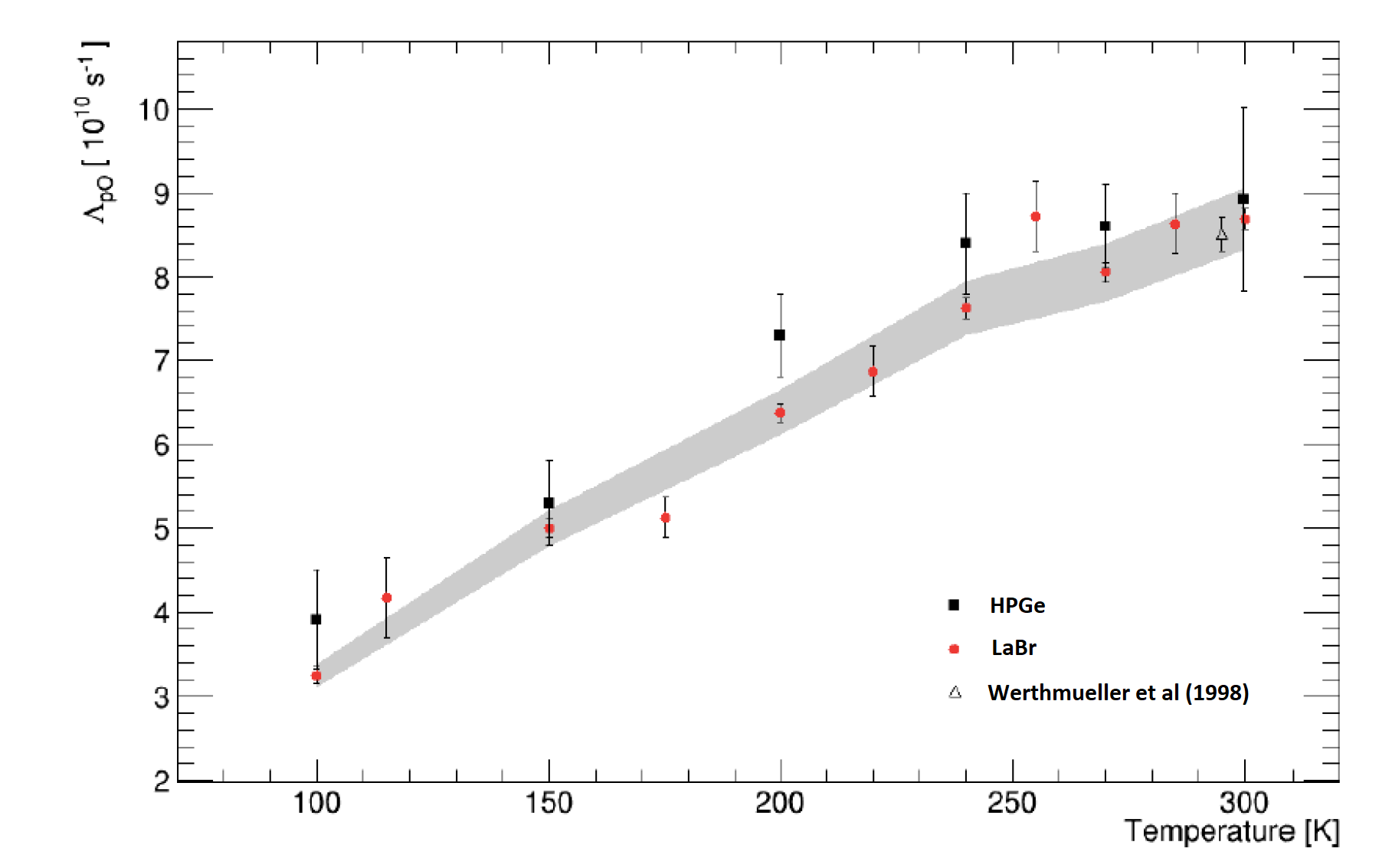}}
\end{center}
\caption{
Temperature dependence of the muon transfer rate to oxygen as measured by FAMU independently with HpGe and LaBr detectors. Error bars indicate the statistical 
error only. The shaded region represents the systematic uncertainty of the 
result with LaBr$_3$:Ce. Data taken at PSI \cite{36wertmueller} are also
reported.}
\label{transfer}
\end{figure}

The efficiency of the method proposed by the FAMU collaboration 
relies on the muon transfer 
rate dependence on the muonic hydrogen energy and on the epi-thermality 
of the muonic hydrogen at the moment of the muon transfer. In our  investigation of the temperature dependence of the muon-transfer process from the thermalized $\mu$p atoms to oxygen, a strong monotonic rise of the transfer rate to oxygen in the temperature interval 103$\div$300~K has been observed \cite{38mocchiutti},
confirming previous experimental results at PSI \cite{36wertmueller}.
Results on temperature dependence of the muon transfer rate to oxygen, 
as measured with LaBr$_3$:Ce and HpGe detectors are shown in figure 
\ref{transfer}. Their theoretical interpretation is reported in 
reference \cite{38mocchiutti}, where a full discussion of systematics is also reported.
Such a strong change enables to employ the muon transfer rate to oxygen as a signature of the kinetic-energy gain of the $\mu$p atom.

\section{The final experiment layout and expected performances}

As the probability to stimulate an HFS transition is proportional to
$W/\sqrt(T)$, with W laser pulse energy and T target temperature, 
the temperature should  be the lowest possible to maximize the transition probability. At the same time, since  a gas mixture of 
hydrogen and oxygen is used, the condensation temperature of oxygen limits the lowest possible temperature to about 60 K. 

\begin{figure}
\begin{center}
  \includegraphics[width=.5\textwidth]{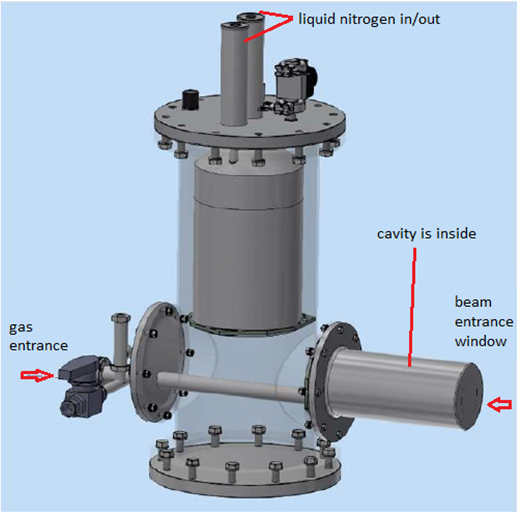}
\end{center}
\caption{Current  design  the cryogenic gas target system.
The optical cavity is not visible and is placed just behind the beam entrance 
window. Muon beam enters from the right.  
}
\label{target}      
\end{figure}

In addition,  to increase the interaction path between the $\mu^{-}$p atoms and the 
photons of the laser pulse an efficient multipass cavity is needed on the 
lines of what previously realized \cite{47vogelsang}. 
The optical cavity design is under study and presently encompass 
a $1.2 \times 1.2 \times 5 $ cm$^3$ illuminated volume in the target
, transverse to the impinging muon beam \cite{baka}.
The optical path leading the laser light towards the internal optical cavity will be sensitive to vibrations. 
Therefore, the target cooling is based on a passive liquid nitrogen system to optimize 
mechanical stability and eliminate the vibrations produced by the cryogenic 
pumps. Figure~\ref{target} shows the actual target design, which has led to the choice of a liquid nitrogen (LN$_2$) 5 liters reservoir, 
that will allow a duty cycle of about 5 days at the operating temperature 
of 80 K. 

The pulsed, tunable FAMU laser at 6785 nm has target parameters of
an energy output of more than 1 mJ, with a line width less 
than 0.07 nm \footnote{pulse duration 10 ns at a 25 Hz repetition rate}. 

The chosen scheme is based on direct frequency generation (DFG) in LiInS2 
crystals with pump and signal coming from one narrowband fixed wavelength 
Nd:Yag laser (1064 nm) and a tunable narrowband Cr:forsterite laser 
(1262 nm) pumped by another Nd:Yag laser, synchronized to the first one.
For the generation of an energy greater than 1 mJ at $7\mu$m the Nd:Yag
laser must have an energy around 100 mJ and the Cr:forsterite of 35 mJ.
For details on the layout of the laser system see reference \cite{stoychev}.  
The laser wavelength ($\sim 6785$ nm) will be measured precisely  
($\sim 30$ ppm) with  solid state Fizeau interferometers and a cell 
filled with $^{12}C_{2}H_{4}$. 


From a detailed Montecarlo simulation the signal to background ratio may
be estimated, optimizing the data taking conditions. For what concerns the
target the optimal conditions require a temperature of 80 K, at a pressure of
7 bar for a $1 \%$ oxygen mixture. With the final laser energy of $\sim 4$ 
mJ we expect a signal to background ratio around 10 in the selected time
window. Background will be evaluated taking alternatively 
one beam spill with laser on 
and one with laser off.

\section{Conclusions}
The method proposed by the FAMU collaboration to measure the Zemach proton 
radius has been demonstrated feasible via our measurement of the transfer 
function to oxygen. The full experimental setup is under realization,
including the optical cavity, the target and the DFG MIR laser. 
A first spectroscopic run is foreseen for the beginning of 2020, before
the long ISIS 2021 shutdown. 

\end{document}